# Coverage and Connectivity in Three-Dimensional Networks

S. M. Nazrul Alam
Department of Computer Science
Cornell University
Ithaca NY 14853 USA
smna@cs.cornell.edu

Zygmunt J. Haas
Department of Electrical and Computer Engineering
Cornell University
Ithaca NY 14853 USA
haas@ece.cornell.edu

http://wnl.ece.cornell.edu

## ABSTRACT
Although most wireless terrestrial networks are based on two-dimensional (2D) design, in reality, such networks operate in three-dimensions (3D). Since most often the size (i.e., the length and the width) of such terrestrial networks is significantly larger than the differences in the third dimension (i.e., the height) of the nodes, the 2D assumption is somewhat justified and usually it does not lead to major inaccuracies. However, in some environments, this is not the case; the underwater, atmospheric, or space communications being such apparent examples. In fact, recent interest in underwater acoustic ad hoc and sensor networks hints at the need to understand how to design networks in 3D. Unfortunately, the design of 3D networks is surprisingly more difficult than the design of 2D networks. For example, proofs of Kelvin's conjecture and Kepler's conjecture required centuries of research to achieve breakthroughs, whereas their 2D counterparts are trivial to solve. In this paper, we consider the coverage and connectivity issues of 3D networks, where the goal is to find a node placement strategy with 100% sensing coverage of a 3D space, while minimizing the number of nodes required for surveillance. Our results indicate that the use of the Voronoi tessellation of 3D space to create truncated octahedral cells results in the best strategy. In this truncated octahedron placement strategy, the transmission range must be at least 1.7889 times the sensing range in order to maintain connectivity among nodes. If the transmission range is between 1.4142 and 1.7889 times the sensing range, then a hexagonal prism placement strategy or a rhombic dodecahedron placement strategy should be used. Although the required number of nodes in the hexagonal prism and the rhombic dodecahedron placement strategies is the same, this number is 43.25% higher than the number of nodes required by the truncated octahedron placement strategy. We verify by simulation that our placement strategies indeed guarantee ubiquitous coverage. We believe that our approach and our results presented in this paper could be used for extending the processes of 2D network design to 3D networks.

## Categories and Subject Descriptors
C.2.1 [**Computer-Communication Networks**]: Network Architecture and Design – *network topology*, *wireless communication*.

## 1. INTRODUCTION
Design of terrestrial networks is generally two-dimensional where it is assumed that all nodes of a network reside on a plane. This assumption may no longer be valid if a network is deployed in space, atmosphere, or ocean, where nodes of a network are distributed over a 3D space. Although such a scenario may not be common at present, applications are being developed that will make three-dimensional networks increasingly common in the near future. For example, recently underwater acoustic ad hoc and sensor networks have generated a lot of interest among the researchers [1], [12], [15], [27]. Ocean column monitoring requires the nodes to be placed at different depths of the water, thus creating a three-dimensional network [1]. Weather forecasting and climate monitoring can also benefit if three-dimensional networks can be deployed in the atmosphere.

In this paper, we focus on the coverage and connectivity issues of three-dimensional networks, where all the node have the same sensing range and the same transmission range. In particular, we want to answer the following questions:

- What is the best way to place the nodes in three-dimension such that the number of nodes required for surveillance of a 3D space is minimized, while guaranteeing 100% coverage?
- What should be the minimum ratio of the transmission range and the sensing range of such a placement strategy?

The answers to these questions have both civilian and military applications. For example, one can envision a scenario where the Air Force uses unmanned aerial vehicles with limited sensing range to form a 3D network for surveillance of an airspace. Similarly, the Navy can use a 3D network of underwater autonomous vehicles for surveillance of ocean. In either case, it is always desirable to find the optimal placement of vehicles in three-dimensions, such that the number of vehicles required is minimized, while 100% coverage in 3D is guaranteed. Similarly, civilian applications include environment and climate monitoring in ocean and atmosphere using ad hoc and sensor networks.

Proving optimality in many 3D problems is surprisingly difficult, even though similar problems in 2D can be proved easily. For example, Kepler's sphere packing problem has been around since 1611 and a proof of Kepler's conjecture has only been found in 1998. Similarly, no optimality proof for Kelvin's 1887 conjecture is known yet. In this paper, we show similarity between our problem and Kelvin's problem. We use Kelvin's conjecture to justify that the placement of nodes in the middle of truncated octahedrons cells, which are, created by Voronoi tessellation of a 3D space, is the answer to the first question. Like the proof of Kepler's conjecture and Kelvin's conjecture, it is expected that any rigorous proof of our conjecture will be very difficult. However, we provide detailed comparison of this solution with other possible solutions, and we show that this solution is indeed superior to other plausible solutions. Our contributions, results, and conclusions of this paper can be summarized as follows:

- Using Kelvin's conjecture, we conjecture that the answer to our first question is the placement strategy that places nodes in the middle of truncated octahedral cells, created by the Voronoi tessellation of a 3D space..
- We define a metric called *volumetric quotient,* which is the ratio of the volume of a polyhedron to the volume of its circumsphere. If Voronoi tessellation of a 3D space gives identical space-filling polyhedron of a fixed radius, then the higher the volumetric quotient of that polyhedron, the smaller the number of nodes required for full 3D coverage.
- We show that the volumetric quotient of truncated octahedron is 0.68329, much higher than other possible space-filling polyhedron. For example, the volumetric quotient of rhombic dodecahedron is 0.477; hexagonal prism has volumetric quotient of 0.477, and cube has just 0.36755. These results imply that a placement strategy whose Voronoi tessellation creates truncated octahedral cells requires 43.25% fewer nodes than the placement strategy whose Voronoi tessellation have rhombic dodecahedron or hexagonal prism as the shape of its cells.
- We show how to place nodes using any of these placement strategies. For each placement strategy, we define a new $u,v,w$-coordinate system, where a node should be placed on each integer coordinate of this new system. Relation of this new $u,v,w$-coordinate system with the original given $x,y,z$-coordinate system has been provided in equations (1), (2), (3) and (4) in terms of the sensing range $R$ and the location of an arbitrary node in the original $x,y,z$-coordinate system ($cx,cy,cz$). Our strategies require only a constant number of arithmetic operations to compute the location of each node and hence is computationally very efficient.
- We find that truncated octahedron placement strategy requires that the ratio of transmission range to the sensing range must be at least 1.7889 in order to maintain connectivity among nodes. For cube, hexagonal prism, and rhombic dodecahedron placement strategies, the ratio is 1.1547, 1.4142, and 1.4142, respectively.

The rest of the paper is organized as follows. Section 2 presents some relevant background information on space-filling polyhedron, Voronoi tessellation, and the famous conjectures of Kelvin and Kepler. The section also presents the related work in the technical literature. Section 3 formally introduces the problem and the assumptions used. In section 4 we analyze the problem and derive the results. Section 5 provides simulation results, which demonstrate that our placement strategies indeed guarantee full coverage. Section 6 discusses the results and proposes future directions of this work. Finally, section 7 concludes the paper.

## 2. PRELIMINARIES

In this section, we define some relevant terms and provide some background information necessary to explain the problem and our approach. The last subsection describes related works in technical literature.

### 2.1 *Space-Filling Polyhedron*

A polyhedron is a three-dimensional shape consisting of finite number of polygonal faces. The faces meet in straight line segments called edges, and the edges meet at points called vertices of the polyhedron. A polyhedron surrounds a bounded volume in three-dimensions. Example of polyhedrons includes cube, prisms, and pyramids. Polygon is a two-dimensional analog of polyhedrons. The general term for any dimension is polytope.

A space-filling polyhedron is a polyhedron that can be used to fill a volume without any overlap or gap (a.k.a. tessellation or tiling). Since the sensing region of a node is spherical and spheres do not tessellate in 3D, we want to find a space-filling polyhedron that best approximates the sphere. In other word, we want to find a space-filling polyhedron such that if each cell is modeled by that polyhedron, then the number of cells required to cover a volume is minimized, where the distance from the center of a cell to its farthest corner (i.e. radius of a cell) is not greater than the sensing range $R$.

At first, we provide a short overview on space-filling polyhedron. It is not easy to show that a polyhedron has space-filling property. For example, although Aristotle claimed that the tetrahedron fills space [2], his claim was incorrect [11], and the mistake remained unnoticed until the 16[th] century [17].

Some of the important results on space-filling polyhedron are as follows: There are exactly five regular polyhedrons (a.k.a. platonic solids or regular solids) [23]: cube, dodecahedron, icosahedron, octahedron, and tetrahedron, as was proved by Euclid in the last proposition of the *Elements*. Cube is the only space-filling regular polyhedron [9]. There are only five convex polyhedrons with regular faces having space-filling property: triangular prism, hexagonal prism, cube, truncated octahedron [23] [29], and gyrobifastigium [13]. The rhombic dodecahedron, elongated dodecahedron, and squashed dodecahedron are also space-fillers. A combination of tetrahedrons and octahedrons fills space. In addition, octahedrons, truncated octahedrons, and cubes, combined in the ratio 1:1:3, can also fill space.

However, we impose the restriction that the shape of the Voronoi cells should be identical, i.e., only one type of polyhedron is used to fill the space. We have two reasons for this requirement:

- algorithms, especially distributed algorithms, to find the location of nodes are far simpler when one type of polyhedron is used, and
- since the radius of the polyhedron is fixed, it is unlikely that any significant improvement can be achieved by using two or more type of polyhedrons to fill the space.

## 2.2 Kelvin's Conjecture

Now we describe the century old Kelvin's conjecture that we will use in section 4 to justify why truncated octahedron is the most likely building block for the optimal solution.

In 1887, Lord Kelvin asked the following question [25]: *"What is the optimal way to fill a three dimensional space with cells of equal volume, so that the surface area (interface area) is minimized?"* This is essentially the problem of finding a space-filling structure having the highest *isoperimetric quotient*. If the volume and surface area of a structure are $V$ and $S$, respectively, then in three-dimensions its isoperimetric quotient can be defined as $36\pi V^2/S^3$. Sphere has the highest isoperimetric quotient and it is 1. Kelvin's answer to his question was 14-sided truncated octahedron having a very slight curvature of the hexagonal faces and its isoperimetric quotient is 0.757. But Kelvin couldn't prove that the structure is optimal. The uncurved truncated octahedron has isoperimetric quotient of 0.753367. For more than a century, Kelvin's solution was generally accepted as correct [30] and has been widely known as *Kelvin's conjecture*. But in 1994, two physicists Denis Weaire and Robert Phelan came up with another space-filling structure. It consists of six 14-sided polyhedrons and two 12-sided polyhedrons with irregular faces of equal volume that has 0.3% less surface area than truncated octahedron [31], [28]. The isoperimetric quotient of this structure is 0.764. But any proof that the structure of Weiare and Phelan is optimal or that Kelvin's solution is optimal for the identical cells case is yet to be found.

## 2.3 Voronoi Tessellation

In three-dimension, for any (topologically) discrete set $S$ of points in Euclidean space, the set of all points closer to a point $c$ of $S$ than to any other point of $S$ is the interior of a convex polyhedron called the *Voronoi cell* of c. The set of such polyhedrons tessellate the whole space, and is called the *Voronoi tessellation* corresponding to the set $S$. Voronoi tessellation of any solution to our problem of the optimal location of the nodes, gives the optimal shape of each cell.

## 2.4 Kepler's Conjecture

Another closely related problem is Kepler's sphere packing problem. The problem is to find the most efficient way to pack equal-sized spheres. In 1611, Kepler made a guess that the face-centered cubic (FCC) lattice was the most efficient of all arrangements, but was unable to prove it. After four hundred years of failed efforts, Kepler's conjecture was finally proved to be correct by Thomas Hales in 1998 [10]. The proof extensively uses methods from the theory of global optimization, linear programming, and interval arithmetic. The computer code and data files used for the proof required more than 3 gigabytes of space for storage. The Voronoi tessellation of the FCC lattice is rhombic dodecahedron. Although FCC lattice is the optimal solution for sphere packing, in this paper we will show that truncated octahedron, which is the Voronoi tessellation of body-centred cubic (BCC) lattice, actually require 43.25% fewer nodes for our problem. This significant difference is not very intuitive. Note that the FCC lattice has packing density of 74.048% (optimal solution for sphere packing), while BCC lattice has packing density of about 68%.

## 2.5 Related Works in Technical Literature

The problem of finding appropriate locations of base stations in two-dimensional plane, such that the number of base stations required is minimized while 100% coverage is guaranteed, has been solved for cellular networks [22]. In two-dimensional cellular systems the cells are modeled as regular hexagons, such that the radius of each hexagon is equal to the maximum range of a base station.

In sensor network, the problem of providing sensing coverage has received significant attention in the context of two-dimensional networks. Maximizing the sensing coverage is a fundamental requirement for many critical applications of sensor networks e.g., detection [26], monitoring, and tracking and classification [19]. Every point of a selected geographic region must be within the sensing range of at least one sensor in a fully covered network. Several algorithms [7], [5], [21], [36] have been proposed to achieve full sensing coverage in two-dimensional network when a sensor network is deployed using random network topology. In an attempt to maximize the lifetime of a sensor network, energy conservation protocols [24], [32], [34], [35] dynamically maintain sensing coverage by keeping active only a subset of nodes at any particular time. Impact of sensing coverage on the performance of greedy geographic routing has been studied in [33] for 2D wireless sensor networks.

There exist only few references on three-dimensional networks in the literature; the works presented in [6] and [8] studied 3D cellular networks. In [6] each cell is represented as rhombic dodecahedron and in [8] each cell is represented as hexagonal prism. However, in this paper, we show that if truncated octahedron is used to model the shape of a cell, then the required number of nodes to monitor a 3D space is 43.25% fewer than the case where the cell is represented as hexagonal prism or rhombic dodecahedron.

## 3. PROBLEM STATEMENT

The main assumptions and the goals of our work are defined as follows:

### 3.1 *Assumptions of the work*
- All nodes have identical sensing range $R$. Sensing is omnidirectional and the sensing region of each node can be represented by a sphere of radius $R$, having the sensor node at its center.
- The sensing range $R$ is much smaller than the length, the width, or the height of the 3D space to be covered, so that the boundary effect is negligible and hence can be ignored.
- Any point in the 3D space to be covered must be within the sensing range $R$ from at least one node.
- If the locations of the nodes are fixed, their location is arbitraty. If the nodes are mobile, the nodes are initially randomly deployed, and their movement is unrestricted. Thus, we ignore the physical constraints of placing the nodes, and we assume that the placement strategy is free to place a node at any location in the network.

### 3.2 *Goals of the work*
- Given any $R$, find the number of nodes and their locations, such that the number of nodes required to cover any specified volume is minimized.
- For placement strategies that achieve the above goal, find the minimum transmission range in terms of sensing range $R$, such that the all nodes are connected to their neighbors.

## 4. ANALYSIS

In this section, we analyze our problem from the point of view of the shape of Voronoi cells corresponding to the placement of nodes in the network. If each Voronoi cell is identical and the boundary effect neglected, then total number of nodes required for 3D coverage is simply the ratio of the volume of the 3D space to be covered to the volume of one Voronoi cell. So minimizing the number of nodes can be achieved if the Voronoi cells have the highest volume for the given sensing radius $R$. Clearly, the radius of the circumsphere of a Voronoi cell must be less than or equal to the sensing range $R$. Since achieving the highest volume is the goal, the radius of circumsphere must always be equal to the sensing range $R$. And since $R$ is fixed, the volumes of the circumspheres of all Voronoi cells are the same and equal to $4\pi R^3/3$. Finally, the shape of any Voronoi cell in 3D is always a polyhedron. The restriction of identical Voronoi cell implies that the polyhedron must also have the space-filling property. So our problem reduces to the problem of finding the space-filling polyhedron which has the highest ratio of its volume to the volume of its circumsphere. We call this ratio the *volumetric quotient* of the space-filling polyhedron. More formal definition is as follows:

**Definition:** *For any polyhedron, if the maximum distance from its center to any vertex is R and the volume of that polyhedron is V, then the volumetric quotient of that polyhedron is*

$$\frac{V}{\frac{4}{3}\pi R^3}.$$

Since the volume of the circumsphere is the upper bound on the volume of any polyhedron, the value of volumetric quotient is always between 0 and 1. Clearly, for a given sensing range $R$, the number of nodes required to cover a 3D space is inversely proportional to the volumetric quotient of the space-filling polyhedron used as a Voronoi cell. So our problem reduces to the problem of finding the space-filling polyhedron that has the highest volumetric quotient.

Finding the optimal polyhedron and proving its optimality seems to be very hard; indeed, some problems in 3D that have optimality criterion associated with them took centuries to prove (Kelvin's problem is still open after more than one century, while Kepler's conjecture was proven only recently after almost five centuries of efforts). Since providing any rigorous proof is likely to be an intractable problem, we proceed in the following way. At first we provide some intuition why truncated octahedron is the most likely solution by drawing similarity between our problem and Kelvin's conjecture. Then we choose three other different space-filling polyhedrons that have been used by other researchers in similar problems and which are reasonable contenders to the truncated octahedron. We provide detailed comparison among these four space-filling polyhedrons and we show that truncated octahedron has much higher volumetric quotient than others. Thus truncated octahedron requires much fewer nodes than the other shapes for coverage in a 3D network. We also provide the placement strategies that create Voronoi cells having shape of our chosen polyhedron. Finally, the connectivity issue has been addressed by determining the minimum transmission radius needed to maintain connectivity among neighboring nodes in various placement strategies.

### 4.1 *Similarity with Kelvin's Conjecture*
Kelvin's problem is essentially finding a space-filling polyhedron that has minimum ratio of surface-area to volume. We claim that the space-filling polyhedron with the minimum ratio of surface-area to volume best approximates the sphere. It is well known that among all structures, the following claims hold:

 1. For a given volume, sphere has the smallest surface area.

2. For a given surface area, sphere has the largest volume.

From the above two statements, we claim the following. Suppose that any two space-filling polyhedrons $P_1$ and $P_2$ have equal volume. If surface-area of $P_1$ is smaller than the surface area of $P_2$, then $P_1$ is a better approximation of a sphere than $P_2$. Again if $P_1$ is a better approximation of a sphere than $P_2$, then $P_1$ has higher volumetric quotient than $P_2$. Recall that among all shapes, sphere has the highest volumetric quotient and it is 1.

So the solution to Kelvin's problem is essentially the solution to our problem. Since until now, truncated octahedron is the best solution for Kelvin's problem for the case of a single cell shape, we conjecture that truncated octahedron is the most likely solution to our problem as well. Note that, we will consider the uncurved version of the truncated octahedron because it is mathematically more tractable than the curved version and the difference between the curved version and the uncurved version is negligible. Because the argument given above is not rigorous enough, to increase the confidence in our solution, we choose other likely contenders to the truncated octahedron, and we provide comparison of the truncated octahedron with those other space-filling polyhedrons.

## 4.2 Choice of Other Polyhedrons

One can try to solve our problem using Kepler's problem in the following way. Find the maximal packing of spheres and then take the Voronoi tessellation corresponding to the centers of the spheres. Take the radius of spheres such that the maximum distance from a center to any vertex of the corresponding Voronoi cell is the sensing range $R$. Kepler's conjecture for sphere packing problem has been proven recently after five centuries of efforts, with the FCC lattice being the solution to that problem. The Voronoi tessellation of the FCC lattice is rhombic dodecahedron. So we choose rhombic dodecahedron as one of the contender to the truncated octahedron.

The solution to our problem in 2D is the hexagon. The polyhedron that has hexagon as its cross section in all the three directions ($x$, $y$, and $z$) does not have space-filling property. The polyhedrons that have space-filling property and hexagonal cross section are rhombic dodecahedron and hexagonal prism. So we include hexagonal prism in our comparison as well. Finally, most simplistic choice is the cube and it is the only regular polyhedron that tessellates in 3D space. So we compare the truncated octahedron with the rhombic dodecahedron, the hexagonal prism, and the cube, and we show that the truncated octahedron has better volumetric quotient that the other choices, hence requiring fewer nodes. A rigorous proof that considers all possible space-filling polyhedrons are intractable, as is evident from the fact that Kelvin's problem for the case of a single cell shape is still open for more than a century.

## 4.3 Volumetric Quotient and Number of Nodes Required by Different Polyhedrons

Here we calculate the volumetric quotients of our chosen polyhedrons and also provide a comparison of the number of nodes required when each of the polyhedrons is used as Voronoi cells.

### 4.3.1 Cube

If the length of each side of a cube is $a$, then the radius of its circumsphere is $\sqrt{3}a/2$. So the volumetric quotient of a cube is: $a^3 \left/ \frac{4}{3}\pi\left(\frac{\sqrt{3}}{2}a\right)^3 \right. = \frac{2}{\sqrt{3}\pi} = 0.36755$.

### 4.3.2 Hexagonal Prism

The volumetric quotient of a hexagonal prism depends on its height. So at first we need to find out the optimal height of a hexagonal prism which has the largest volumetric quotient among all the hexagonal prisms. Suppose that the length of each side of the hexagon is $a$ and the height of the hexagonal prism is $h$. Then the radius of circumsphere of the hexagonal prism is $\sqrt{a^2 + h^2/4}$, the volume of the hexagonal prism is $3\sqrt{3}a^2h/2$ and the volumetric quotient is 
$$\frac{3\sqrt{3}}{2}a^2h \left/ \frac{4}{3}\pi\left(\sqrt{a^2+\frac{h^2}{4}}\right)^3 \right. .$$

If we set the first derivative of the volumetric quotient to zero, then we obtain that

$$\frac{\frac{3\sqrt{3}}{2}a^2}{\frac{4}{3}\pi\left(\sqrt{a^2+\frac{h^2}{4}}\right)^3} - \frac{3}{2}\frac{\frac{3\sqrt{3}}{2}a^2h\cdot\frac{2h}{4}}{\frac{4}{3}\pi\left(\sqrt{a^2+\frac{h^2}{4}}\right)^5} = 0$$

$$\Rightarrow a^2 + h^2/4 = 3h^2/4$$

So, the optimum value of $h$ is $a\sqrt{2}$ and the optimum volumetric quotient for hexagonal prism is

$$\frac{3\sqrt{3}}{2}a^2 a\sqrt{2} \left/ \frac{4}{3}\pi\left(\sqrt{a^2+\frac{a^2}{2}}\right)^3 \right. = \frac{6}{4\pi} = 0.477.$$

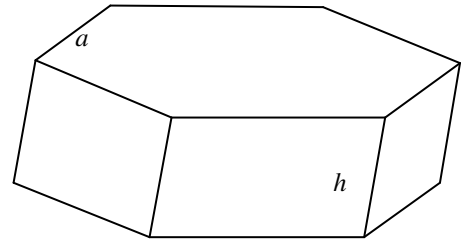

**Figure 1. A Hexagonal Prism**

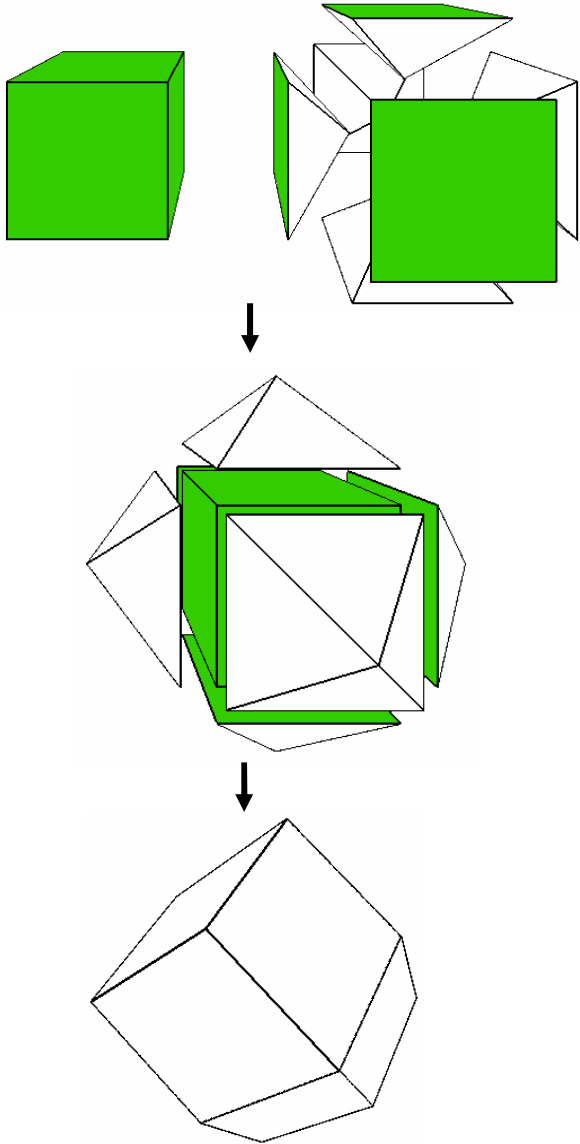

**Figure 2. Construction of a rhombic dodecahedron from two identical cubes**

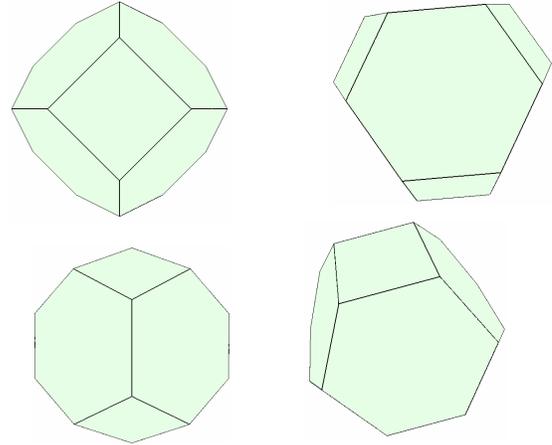

**Figure 3. Truncated Octahedron.**

*4.3.3    Rhombic Dodecahedron*

A rhombic dodecahedron can be constructed from two identical cubes in the following way, shown in Figure 2. Take one cube and cut it into six equal pyramids such that the base of each pyramid consists of one face of the cube. Take another similar cube and place each pyramid on the cube, such that the base of the pyramid is on one side of the cube. This creates a rhombic dodecahedron. If each side of the two original cubes is *a* (i.e., length of each edge of the rhombic dodecahedron is $\sqrt{3}a/2$), then total volume of the rhombic dodecahedron is $2a^3$. The center of the rhombic dodecahedron is the center of the second (intact) cube. Eight vertices of the intact cube form eight vertices of the rhombic dodecahedron and their distance from the center is the radius of the circumsphere of the cube, equal to $\sqrt{3}a/2$. The other six vertices of the rhombic dodecahedron are formed by the six pieces of the first cube. Distance from the center of second cube to its surface is $a/2$ and the height of each of the six pyramids is $a/2$. So the distance from the center of rhombic dodecahedron to each of these six vertices is *a*, and this is also the radius of the circumsphere of the rhombic dodecahedron. So the volumetric quotient of rhombic dodecahedron is $2a^3 / \frac{4}{3}\pi a^3 = 6/4\pi = 0.477$, which is exactly the same as that of the hexagonal prism.

*4.3.4    Truncated Octahedron*

Truncated octahedron has 14 faces, of which 8 are hexagonal and 6 are square faces, and the length of the edges of hexagons and squares are the same. Suppose that the length of each edge is *a*. The distance between two opposite hexagonal faces is $\sqrt{6}a$ and the distance between two opposite square faces is $2\sqrt{2}a$. The radius of the circumsphere of the truncated octahedron is $\sqrt{10}a/2$. The volume of the truncated octahedron is $8\sqrt{2}a^3$ and the volumetric quotient is

$$8\sqrt{2}a^3 / \frac{4}{3}\pi\left(\frac{1}{2}\sqrt{10}a\right)^3 = 24/5\sqrt{5}\pi = 0.68329.$$

*4.3.5    Comparison*

Among all the polyhedrons considered, the truncated octahedron gives the best volumetric quotient. We can also compare the number of nodes required by each type of polyhedron. The number of nodes required by the cube is 0.68329/0.36755=1.859 times that of the truncated octahedron. For the hexagonal prism this value is 0.68329/0.477=1.4325 and for the rhombic dodecahedron it is 0.68329/0.477=1.4325. Table I summarizes the results.

It is interesting to see how this result relates to the 2D networks. Hexagon has the optimal tiling in 2D. The ratio of the area of a hexagon and the area of its circumcircle is $3\sqrt{3}/2\pi = 0.82699$. It is not difficult to see why the quotient in 3D is lower than that of 2D. In 1D tiling, we can achieve quotient of 1 by using straight line tiling (actually there is only one possible tiling in 1D). One can observe that the proportional loss in the value of quotient remains roughly the same as we go to the higher dimensions. If we assume that the truncated octahedron is indeed the best shape for 3D tiling, then its quotient is 82.623% (0.68329/0.82699=0.82623) of the quotient achieved by hexagon in 2D, which is again 82.699% of the quotient of 1D tiling.

**Table I: Volumetric Quotient of Different Types of Space-filling Polyhedrons**

| Polyhedron | Volumetric quotient | Number of nodes needed Compared to truncated octahedron |
|---|---|---|
| Cube | 0.36755 | 85.9% more |
| Hexagonal Prism | 0.477 | 43.25% more |
| Rhombic Dodecahedron | 0.477 | 43.25% more |
| Truncated Octahedron | 0.68329 | same |

### 4.3.6 Explanation of Why Truncated Octahedron is Better

Cross sections of the rhombic dodecahedron and the hexagonal prism are hexagons, but the vertices of this hexagon are not on the great circle of the circumsphere. As a result, the radius of the hexagon is smaller than the sensing range. In the case of truncated octahedron, the two dimensional tiling is by octagons that lies on the great circle. But 2D tiling by regular octagon has square gaps (See Fig. 4). These square gaps are filled in 3D by cells from one level above and one level below. As a result, for a given sensing radius, truncated octahedron requires fewer cells to cover a given 3D space.

## 4.4 Placement Strategies

In this subsection, we provide strategies (algorithms) to pinpoint the location where the nodes should be placed such that the Voronoi cells are our chosen space-filling polyhedrons. We take an arbitrary point (preferably the center of the space to be covered) as an input and place a node there. Then we find the locations of other nodes relative to this center node. So the input to our algorithm is sensing range $R$ and the co-ordinates of the point, say ($cx, cy, cz$), which act as a seed for the growing lattice. Distributed versions of the algorithms can use the location of the leader node as the seed.

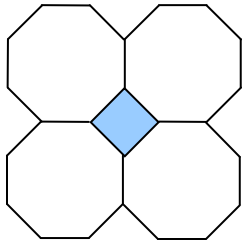

**Figure 4. Octagon does not tile a plane.**

### 4.4.1 Cube

In each direction parallel to the $x$, $y$, and $z$ axes, the distance between any two neighboring nodes is $2R/\sqrt{3}$. If a node is placed on each integer coordinate of the following coordinate system then we obtain cube tessellation. Suppose that the coordinate system is defined by three axes: $u$, $v$, and $w$, which are parallel to the $x$, $y$, and $z$ axes, respectively. The center of the coordinate system is ($cx, cy, cz$) and unit distance in each axes is $2R/\sqrt{3}$. So a node at ($u_1, v_1, w_1$) in the new co-ordinate system should be placed in the original $x,y,z$-coordinate system at

$$\left(cx + u_1 \times \frac{2R}{\sqrt{3}}, cy + v_1 \times \frac{2R}{\sqrt{3}}, cz + w_1 \times \frac{2R}{\sqrt{3}}\right). \quad (1)$$

In general, real distance $d_{12}^{cb}$ between any two points with coordinates $(u_1, v_1, w_1)$ and $(u_2, v_2, w_2)$ in the $u,v,w$-coordinate system is

$$d_{12}^{cb} = \frac{2}{\sqrt{3}} R \sqrt{(u_2 - u_1)^2 + (v_2 - v_1)^2 + (w_2 - w_1)^2}.$$

In order to illustrate how above information can be used to place nodes efficiently, suppose that we want to cover a volume $\left(\frac{100\sqrt{3}}{2R} \times \frac{100\sqrt{3}}{2R} \times \frac{100\sqrt{3}}{2R}\right)$ with a center at $(cx, cy, cz)$. Then the following pseudo-code segment could be used:

```
For u=-50 to 50 do
  For v=-50 to 50 do
    For w= -50 to 50 do
      put_node_at (cx+u×2R/√3, cy+v×2R/√3, cz+w×2R/√3)
```

To save space, from now on, rather than providing formal pseudo-code for all polyhedrons, we only provide all necessary information about the node placement.

### 4.4.2 Hexagonal Prism

Suppose that the hexagons are parallel to the $xy$ plane. Then the distance between neighboring nodes along the $z$-axis is $2R/\sqrt{3}$. Note that the optimal height of the hexagonal prism is $\sqrt{2}$ times the radius of the hexagon. If a node is placed at every integer coordinate of the following coordinate system then we have hexagonal prism tessellation. The axis $v$ is parallel to the axis $y$. The angle between the $u$ and the $v$ axes is $60^0$ in the positive half, and the unit distance along each axis is equal to $R\sqrt{2}$. So the angle between the axis $u$ and the axis $x$ is $30^0$, and the angle between the axis $u$ and the axis $y$ is $60^0$. The $w$ axis is orthogonal to the $uv$ plane and the unit distance along the $w$ axis is $2R/\sqrt{3}$. Thus the axis $w$ is parallel to the axis $z$. Finally, the center of the $u,v,w$-coordinate system is at $(cx, cy, cz)$ point of the $x,y,z$-coordinate system.

A node at $(u_1, v_1, w_1)$ in the new $u,v,w$-coordinate system should be placed in the original $x,y,z$-coordinate system at

$$\begin{pmatrix} cx + u_1 \times R\sqrt{2} \sin 60^0, \\ cy + u_1 \times R\sqrt{2} \cos 60^0 + v_1 \times R\sqrt{2}, cz + w_1 \times \frac{2R}{\sqrt{3}} \end{pmatrix}$$

$$= \left( cx + u_1 R\sqrt{\frac{3}{2}}, cy + (u_1 + 2v_1)\frac{R}{\sqrt{2}}, cz + \frac{2Rw_1}{\sqrt{3}} \right). \quad (2)$$

Real distance between two points with coordinates $(u_1, v_1, w_1)$ and $(u_2, v_2, w_2)$ in the $uvw$-coordinate system is

$$d_{12}^{hp} = R\sqrt{2} \sqrt{\begin{array}{l}(u_2 - u_1)^2 + (u_2 - u_1)(v_2 - v_1) + \\ (v_2 - v_1)^2 + \frac{2}{3}(w_2 - w_1)^2 \end{array}}.$$

### 4.4.3 Rhombic Dodecahedron

If a node is placed at every integer coordinate of the following $u,v,w$ coordinate system then we get rhombic dodecahedron tessellation. The axes $u$ and $v$ are parallel to the axes $x$ and $y$, respectively. The angle between the $u$ and the $w$ axis is $60^0$ in the positive half and the angle between the $v$ and the $w$ axes is also $60^0$ in the positive half. The unit distance along each axis is $R\sqrt{2}$. The angle between the $w$ axis and the $z$ axis is $45^0$. Finally, the center of the $u,v,w$-coordinate system is at the $(cx, cy, cz)$ point of the $x,y,z$-coordinate system.

A node at $(u_1, v_1, w_1)$ in the new $u,v,w$-coordinate system should be placed in the original $x,y,z$-coordinate system at

$$\begin{pmatrix} cx + u_1 \times R\sqrt{2} + w_1 \times R\sqrt{2} \cos 60^0, \\ cy + v_1 \times R\sqrt{2} + w_1 \times R\sqrt{2} \cos 60^0, \\ cz + w_1 \times R\sqrt{2} \cos 45^0 \end{pmatrix}$$

$$= \begin{pmatrix} cx + (2u_1 + w_1)\frac{R}{\sqrt{2}}, cy + (2v_1 + w_1)\frac{R}{\sqrt{2}}, \\ cz + w_1 R \end{pmatrix}. \quad (3)$$

The real distance between two points with coordinates $(u_1, v_1, w_1)$ and $(u_2, v_2, w_2)$ in the $u,v,w$-coordinate system is

$$d_{12}^{rd} = R\sqrt{2} \sqrt{\begin{array}{l}(u_2 - u_1)^2 + (v_2 - v_1)^2 + (w_2 - w_1)^2 + \\ (u_2 - u_1)(w_2 - w_1) + (v_2 - v_1)(w_2 - w_1)\end{array}}.$$

Note that in the case of space-filling by identical rhombic dodecahedron, the distance between the centers of any two neighboring rhombic dodecahedrons is the same. However, this is not the case for the hexagonal prism or the truncated octahedron.

### 4.4.4 Truncated Octahedron

If a node is placed at every integer coordinate of the following $u,v,w$-coordinate system then we get truncated octahedron tessellation. The center of $u,v,w$-coordinate system is at the $(cx, cy, cz)$ point of the $x,y,z$-coordinate system. The axes $u$ and $v$ are parallel to the axes $x$ and $y$, respectively. The unit distance in both the $u$ and $v$ axes is $4R/\sqrt{5}$. The $w$ axis is such that $\angle uv = \angle vw = \cos^{-1}(\sqrt{1/3}) = 54.73^0$ in the positive quadrant and the unit distance in $w$ direction is $2\sqrt{3}R/\sqrt{5}$. The axis $w$ creates an angle of $\cos^{-1}(\sqrt{1/3}) = 54.73^0$ with the $z$ axis.

A node at $(u_1, v_1, w_1)$ in the new $u,v,w$-coordinate system should be placed in the original $x,y,z$-coordinate system at

$$\begin{pmatrix} cx + u_1 \frac{4R}{\sqrt{5}} + w_1 \frac{2\sqrt{3}R}{\sqrt{5}} \cos \alpha, \\ cy + v_1 \frac{4R}{\sqrt{5}} + w_1 \frac{2\sqrt{3}R}{\sqrt{5}} \cos \alpha, \\ cz + w_1 \frac{2\sqrt{3}R}{\sqrt{5}} \cos \alpha \end{pmatrix}, \text{ where } \alpha = \cos^{-1}\left(\sqrt{\frac{1}{3}}\right).$$

After simplifying we obtain

$$\left( cx + (2u_1 + w_1)\frac{2R}{\sqrt{5}}, cy + (2v_1 + w_1)\frac{2R}{\sqrt{5}}, cz + w_1 \frac{2R}{\sqrt{5}} \right) \quad (4)$$

The real distance between two points with coordinates $(u_1, v_1, w_1)$ and $(u_2, v_2, w_2)$ in the $u,v,w$-coordinate system is

$$d_{12}^{to} = \frac{4}{\sqrt{5}} R \sqrt{\begin{array}{l}(u_2 - u_1)^2 + (v_2 - v_1)^2 + \\ (u_2 - u_1)(w_2 - w_1) + \\ (v_2 - v_1)(w_2 - w_1) + \frac{3}{4}(w_2 - w_1)^2 \end{array}}.$$

Location of any node can be found by using equations (1), (2), (3) and (4) for the cube, the hexagonal prism, the rhombic dodecahedron, and the truncated octahedron placement strategies, respectively.

## 4.5 Transmission Range vs. Sensing Range

The required minimum transmission range to maintain connectivity among neighboring nodes depends on the choice of the polyhedron. When the cube is chosen, the distance between two neighboring node is $2R/\sqrt{3}$. Thus the transmission radius must be at least $1.1547R$. If the hexagonal prism is used, then the transmission range must be at least $\sqrt{2}R = 1.4142R$ to maintain connectivity with the neighbors along the axes *u* and *v,* and the transmission range must be at least $2R/\sqrt{3} = 1.1547R$ for communication along the *w*-axis. In the case of the rhombic dodecahedron, the minimum transmission range required is $\sqrt{2}R = 1.4142R$ for communication with any neighbor. Finally, if the truncated octahedron is used, then the transmission range must be at least $4R/\sqrt{5} = 1.7889R$ along the axes *u* and *v*, and the minimum transmission range is $R2\sqrt{3}/\sqrt{5} = 1.5492R$ along the axis *w*. The results are summarized in Table II.

**Table II: Minimum Transmission Range for Different Polyhedrons**

| Polyhedron | Minimum Transmission Range | | | Max of Min Transmission Range |
|---|---|---|---|---|
| | *u*-axis | *v*-axis | *w*-axis | |
| Cube | 1.1547*R* | 1.1547*R* | 1.1547*R* | 1.1547*R* |
| Hexagonal Prism | 1.4142*R* | 1.4142*R* | 1.1547*R* | 1.4142*R* |
| Rhombic Dodecahedron | 1.4142*R* | 1.4142*R* | 1.4142*R* | 1.4142*R* |
| Truncated Octahedron | 1.7889*R* | 1.7889*R* | 1.5492*R* | 1.7889*R* |

## 5. SIMULATION

We wrote our simulation in C using OpenGL and implemented the strategies provided in subsection 4.4. The graphical output shows that placing nodes according to equations (1), (2), (3), and (4) indeed covers the whole space, while the Voronoi cells have corresponding shapes. Our simulation outputs are animation videos from different viewing perspective that gives the viewer a clear understanding of the placement strategies. Unfortunately it is difficult to get a good understanding from still images (i.e., snapshots taken from the animation) as provided in this paper. The program source codes and executable files are available at http://www.cs.cornell.edu/~smna/3DNet/ so that interested readers can download and execute the programs.

Fig. 5 shows the placement of the nodes according to our proposed algorithm based on the truncated octahedron model in a volume where in each dimension the length is 20m and the sensing range is *R*=5m. Each black dot represents a node. A truncated octahedron having radius 5m is drawn around each node in order to show that our placement strategy indeed provides 100% coverage. Axes *x, y,* and *z* are represented by red, green and blue lines, respectively, in the figures to give the reader a perspective about the actual placement. Fig. 6 offers a better view, because it has a smaller number of nodes. These figures show node placement assuming very large network as we ignored the boundary effect. Extra nodes required to fill in the gaps in the boundary may affect the choice of the polyhedron for networks where the sensing range is not very small as compared to the network size. However, for a reasonably large network the boundary effect is negligible.

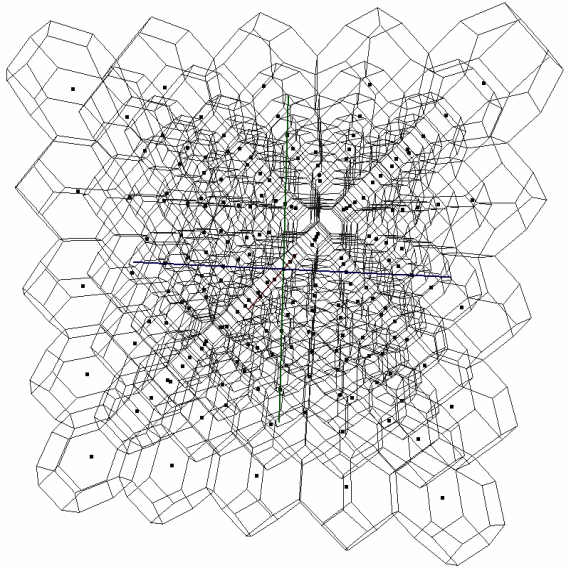

**Figure 5. Truncated Octahedron placement strategy. 3D space is 20mx20mx20m and R=5m.**

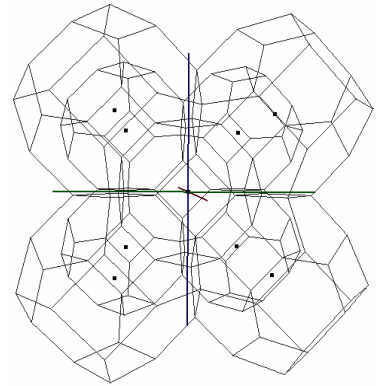

**Figure 6. Truncated Octahedron placement strategy. 3D space is 15mx15mx15m and R=10m.**

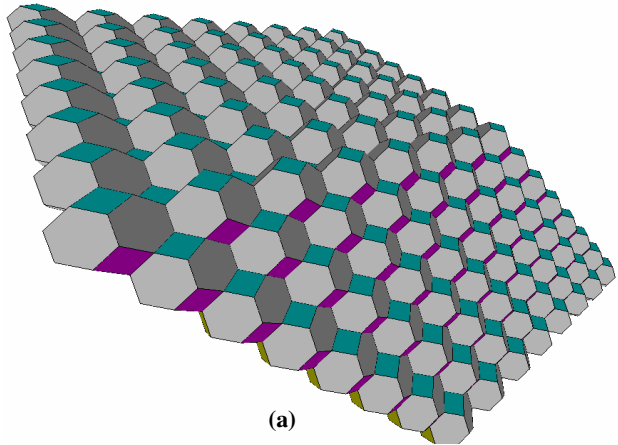

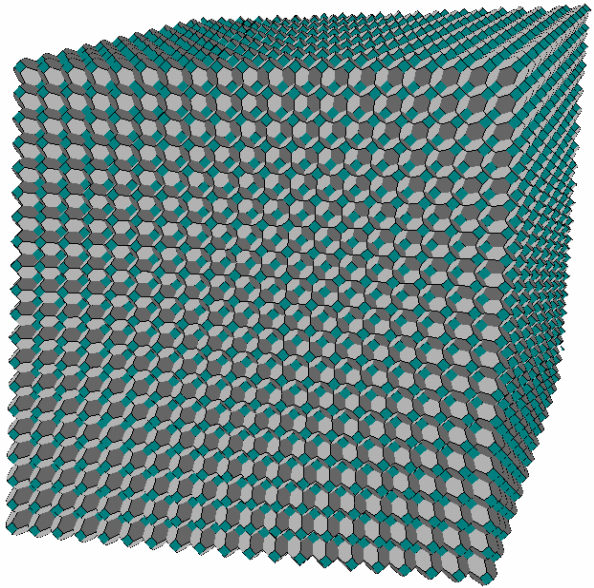

**Figure 7. Boundary of a network consisting of (a) 8x8x8 nodes (b) 20x20x20 nodes when the placement strategy is truncated octahedron. Note that camera positions are different in (a) and (b).**

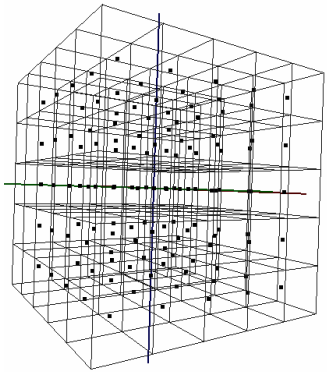

**Figure 9. Node placement based on cube model.**

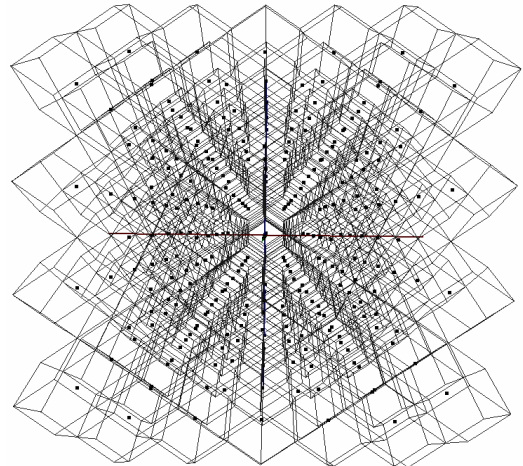

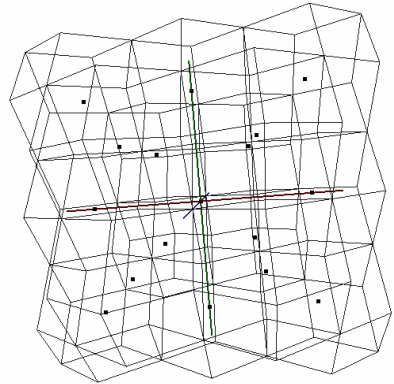

**Figure 8. Rhombic dodecahedron placement strategy. (a) 20mx20mx20m network and $R$=5m (b) 15mx15mx15m network and $R$=10m.**

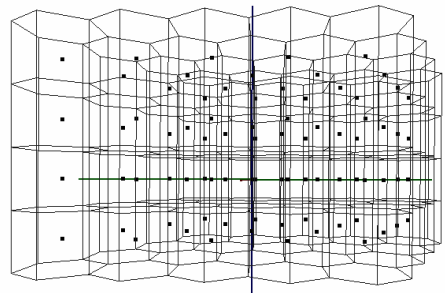

**Figure 10. Hexagonal prism placement strategy.**

Fig. 7 (b) shows that the boundary of a network consisting of 20x20x20 nodes is very much planar. Fig. 8 shows the placement of nodes according to the rhombic dodecahedron model. Placement of nodes according to the cube and the hexagonal prism models are shown in Fig. 9 and Fig. 10, respectively

## 6. DISCUSSION AND FUTURE WORK

This work is applicable to both fixed and mobile network. If the nodes are fixed, then the solution should be used during the initial node deployment. If the nodes are mobile, then the nodes should dynamically compute their desired locations in a distributed fashion and moved to the appropriate location to achieve the above stated goals.

A distributed version of the placement strategies can be devised in the following way. At first, the nodes choose a leader by any standard leader election algorithm [20]. Then the structure can grow relative to the location of the leader. For example, the location of the leader, say ($cx,cy,cz$), can be used as the center of the $u,v,w$-coordinate system and then the same placement strategy described in this paper can be used by other nodes to determine their appropriate location and move accordingly. The placement of nodes can grow like a lattice, using the location of the leader as the seed. This approach works best if the nodes know their location precisely; e.g., using GPS-like system. However, local estimation of distances to neighbors could also be used if the nodes can reach a consensus on the frame of reference ($x,y,z$-axes). For networks in an ocean or in the atmosphere, a good frame of reference can be as follows: the $xy$-plane is parallel to earth surface and the $y$-axis is parallel to the axis going through earth's north-south pole with north-pole in the positive direction. Compass-like instrumentation can be used to find this direction by using earth's magnetic field. Finally, the $z$-axis can be taken as perpendicular to the $xy$-plane with the positive direction away from the earth surface pointing to the space. This approach works if the network is not very big in the $x$- and the $y$-directions, so that the earth surface can be approximated by a plane. A much simpler solutions for this frame of reference problem are likely to exist.

The deterministic approach of finding node placement in this paper assumes that nodes can change location (e.g., surveillance by underwater autonomous vehicles in the ocean or unmanned aerial vehicles in the air). If a node fails to move, other nodes can adjust their locations to fill the gap created by the node failure. However, in the case of 3D space monitoring by inexpensive sensor nodes that cannot adjust their location or don't know their location precisely, many nodes can be randomly distributed in the 3D space. If the network is constructed in the space, there will be no gravitational forces, and the nodes will stay in their initial location with minimum adjustment needed. In the ocean the nodes can be placed in different heights, tethered, to the ocean bottom and using floats to maintain their required depth [1]. In such a network, energy is a major issue and choosing a subset of nodes that remain active at any time while maintaining sensing coverage and connectivity is an important problem. Similar problem in 2D has been addressed in [36]. In 3D, the solution can be found using the truncated octahedron placement strategy provided in this paper as a basis and then by trying to tweak it to accommodate new constrains.

In 2D, routing based on location information has been explored in [18], [3], [4], [16], and [14]. In a 3D network, if placement of the nodes follows any of our placement strategies, then the locations of all other nodes are instantly available from our $u,v,w$-coordinate system. If the $u,v,w$-coordinate of each node is used as its ID, then possible routes between two nodes can be easily determined. Location-based routing protocols can exploit this location-ID information.

## 7. CONCLUSION

In this paper, we investigated the coverage and the connectivity issues in three-dimensional networks, where nodes are placed in a 3D space, unlike most current works that assume the nodes are placed on a 2D plane. Transition from 2D to 3D is not always easy, since many problems in 3D are significantly harder than their 2D counterparts. Many problems that are trivial to solve in 2D turn out to be major research challenges in 3D that remain unsolved even for centuries. In 2D cellular networks, hexagonal tiling is used to place base stations, such that the area covered is maximized with some number of base stations with a fixed radius. The solution of similar problems in 3D is not important for cellular networks since the radius is usually on the order of kilometers. However, the problem is important for other scenarios where nodes with limited sensing range are to cover a vast 3D space. Using century old Kelvin's conjecture, we showed that the truncated octahedral tessellation of 3D space is the most plausible solution for this problem. We defined a metric called *volumetric quotient* that is a measure of the quality of the competing space-filling polyhedrons in our problem. The truncated octahedron turns out to be the best choice with volumetric quotient of 0.68329, which is much better than the volumetric quotients of all the other possible choices (both optimized hexagonal prism and rhombic dodecahedron have volumetric quotient of 0.477, while cube has just 0.36755). The number of nodes required for coverage of a large 3D space depends on the shape of a cell created by the Voronoi tessellation of that space by those nodes. If the shape of each cell is a space-filling polyhedron with higher volumetric quotient, then the number of nodes is smaller. For example, the number of nodes for the rhombic dodecahedron or the hexagonal prism placement requires 43.25% more nodes than the truncated octahedron placement. After finding the optimal placement strategy, we examined the connectivity issues and we found that the best placement strategy (the truncated octahedron) requires the transmission range to be at least 1.7889 times the sensing range in order to maintain full connectivity. We believe that the basic 3D results provided in this paper will be useful in many ways for research and implementation of future three-dimensional networks.